\documentclass[twocolumn,aps,superscriptaddress,showpacs]{revtex4}

\usepackage{amssymb}
\usepackage{amsmath}
\usepackage{graphicx}
\usepackage[normalem]{ulem}
\usepackage[dvips]{color}

\begin{document}

\title{Effects of hadronic potentials on elliptic flows in relativistic heavy ion collisions}
\author{Jun Xu}\email{Jun_Xu@tamu-commerce.edu}
\affiliation{Cyclotron Institute, Texas A\&M University, College
Station, Texas 77843-3366, USA}
\author{Lie-Wen Chen}\email{lwchen@sjtu.edu.cn}
\affiliation{Department of Physics, Shanghai Jiao Tong University,
Shanghai 200240, China}
\author{Che Ming Ko}\email{ko@comp.tamu.edu}
\affiliation{Cyclotron Institute and Department of Physics and
Astronomy, Texas A\&M University, College Station, Texas 77843-3366,
USA}
\author{Zi-Wei Lin}\email{linz@ecu.edu}
\affiliation{Department of Physics, East Carolina University, C-209
Howell Science Complex, Greenville, NC 27858, USA}
\date{\today}

\begin{abstract}

Within the framework of a multiphase transport (AMPT) model that
includes both initial partonic and final hadronic interactions, we
show that including mean-field potentials in the hadronic phase
leads to a splitting of the elliptic flows of particles and their
antiparticles, providing thus a plausible explanation of the
different elliptic flows between $p$ and ${\bar p}$, $K^+$ and
$K^-$, and $\pi^+$ and $\pi^-$ observed in recent Beam Energy Scan
(BES) program at the Relativistic Heavy-Ion Collider (RHIC).

\end{abstract}


\maketitle

Understanding the phase diagram of strongly interacting matters is
one of the main goals of the experiments that are being carried out
in heavy-ion collisions at RHIC~\cite{Bra09}. For collisions at its
highest center-of-mass energy of $\sqrt{s_{NN}}=200$ GeV, convincing
evidences have been established that a strongly interacting
quark-gluon plasma (sQGP) is formed in these
collisions~\cite{Bac05}. The baryon chemical potential of the formed
sQGP is small, and according to calculations based on the lattice
quantum chromodynamics (QCD), the phase transition between the sQGP
and the hadronic matter for such a small baryon chemical potential
is a smooth crossover transition~\cite{Aok06}. Theoretical studies
based on various models have shown, on the other hand, that this
crossover transition would change to a first-order phase transition
when the baryon chemical potential is increased~\cite{Bra09}. To
search for the critical point in the baryon chemical potential and
temperature plane at which such a transition occurs, the BES program
involving Au+Au collisions at lower energies of $\sqrt{s_{NN}}=7.7$,
11.5, and 39 GeV has recently been carried out. Although no
definitive conclusions on the location of the critical point in the
QCD phase diagram have been obtained from these experiments, several
interesting phenomena have been observed~\cite{Moh11}. Among them is
the smaller elliptic flows of ${\bar p}$, $K^-$ and $\pi^+$ than
those of $p$, $K^+$ and $\pi^-$, respectively. The differences
decrease with increasing collision energy and become essentially a
constant at higher collision energies~\cite{Abe07}. These surprising
results were recently attributed to the different elliptic flows of
transported and produced partons during the initial stage of heavy
ion collisions~\cite{Dun11}. Also, it was suggested that the chiral
magnetic effect induced by the strong magnetic field in non-central
collisions could be responsible for the observed difference between
the elliptic flows of $\pi^+$ and $\pi^-$~\cite{Bur11}.

It is known from heavy ion collisions at lower collision energies at
SIS/GSI and AGS/BNL that the elliptic flow of nucleons is affected
not only by their scattering but also by their mean-field potentials
in the hadronic matter~\cite{Dan02}. In particular, in collisions at
a few AGeV when participating nucleons are not blocked by spectator
nucleons, an attractive (repulsive) potential is found to result in
a smaller (larger) elliptic flow. This is because particles with
attractive potentials are more likely to be trapped in the system
and move in the direction perpendicular to the participant plane
while those with repulsive potentials are more likely to leave the
system and move along the participant plane, thus reducing and
enhancing their respective elliptic flows. Also, the potentials of a
particle and its antiparticle are different, and they generally have
opposite signs at high densities~\cite{Ko96,Ko97}. As a result,
particles and antiparticles are expected to have different elliptic
flows in heavy ion collisions when the produced matter has a nonzero
baryon chemical potential. Furthermore, the difference between the
potentials of a particle and its antiparticle diminishes with
decreasing baryon chemical potential, so their elliptic flows are
expected to become similar in higher energy collisions when more
antiparticles are produced. These effects are all consistent with
what were seen in the experimental data from the BES program.

In the present paper, we study quantitatively the elliptic flows of
particles and their antiparticles at BES energies by extending the
AMPT model~\cite{Lin05} to include the potentials of baryons, kaons,
and pions as well as their antiparticles. The AMPT model is a hybrid
model with the initial particle distributions generated by the
Heavy-Ion Jet Interaction Generator (HIJING) model~\cite{Xnw91} via
the Lund string fragmentation model. In the string melting version
of the AMPT model, which is used in the present study, hadrons
produced from excited strings in the HIJING model are converted to
their valence quarks and antiquarks, and their evolution in time and
space is then modeled by Zhang's parton cascade (ZPC)
model~\cite{Zha98}. Different from previous applications of the AMPT
model for heavy ion collisions at higher energies, the parton
scattering cross section and the ending time of the partonic stage
are adjusted in the present study to approximately reproduce
measured elliptic flows and the hadronic energy density ($\sim
0.30-0.35~{\rm GeV/fm^3}$) at the extracted baryon chemical
potential and temperature at chemical freeze out~\cite{And10}.
Specifically, we take the parton scattering cross section to be
isotropic with the value $3$, $6$, and $10$ mb and the ending time
of the partonic stage to be $3.5$, $2.6$, and $2.9$ fm/$c$ for
collisions at $\sqrt{s_{NN}}=7.7$, $11.5$, and $39$ GeV,
respectively. At hadronization, quarks and antiquarks in the AMPT
model are converted to hadrons via a spatial coalescence model, and
the scatterings between hadrons in the hadronic stage are described
by a relativistic transport (ART) model~\cite{Bal95} that has been
extended to also include particle-antiparticle annihilations and
their reverse reactions.

For the nucleon and antinucleon potentials, we take them from the
relativistic mean-field model used in the Relativistic
Vlasov-Uehling-Uhlenbeck transport model~\cite{GQL94}, that is
\begin{eqnarray}
U_{N,{\bar N}}(\rho_B,\rho_{\bar B})&=& \Sigma_s(\rho_B,\rho_{\bar
B}) \pm\Sigma_v^0(\rho_B,\rho_{\bar B}),\label{Up}
\end{eqnarray}
in terms of the nucleon scalar $\Sigma_s(\rho_B,\rho_{\bar B})$ and
vector $\Sigma_v^0(\rho_B,\rho_{\bar B})$ self-energies in a
hadronic matter of baryon density $\rho_B$ and antibaryon density
$\rho_{\bar B}$. The "$+$" and "$-$" signs are for nucleons and
antinucleons, respectively. We note that nucleons and antinucleons
contribute both positively to $\Sigma_s$ but positively and
negatively to $\Sigma_v$, respectively, as a result of the
$G$-parity invariance. Since only the light quarks in baryons and
antibaryons contribute to the scalar and vector self-energies in the
mean-field approach, the potentials of strange baryons and
antibaryons are reduced relative to those of nucleons and
antinucleons according to the ratios of their light quark numbers.

The kaon and antikaon potentials in the nuclear medium are also
taken from Ref.~\cite{GQL94} based on the chiral effective
Lagrangian, that is $U_{K,{\bar K}} = \omega_{K,{\bar K}} -
\omega_0$ with
\begin{eqnarray}
\omega_{K,{\bar K}} &=& \sqrt{m_K^2 + p^2 - a_{K,{\bar K}}\rho_s
+(b_K\rho_B^{\rm net})^2}\pm b_K\rho_B^{\rm net}
\end{eqnarray}
and $\omega_0=\sqrt{m_K^2+p^2}$, where $m_K$ is the kaon mass and
$a_K=0.22$ GeV$^2$fm$^3$, $a_{\bar K}=0.45$ GeV$^2$fm$^3$ and
$b_K=0.33$ GeVfm$^3$ are empirical parameters taken from
Ref.~\cite{GQL97}. In the above, $\rho_s$ is the scalar density,
which can be determined from $\rho_B$ and $\rho_{\bar B}$ through
the effective interaction used for describing the properties of
nuclear matter, and $\rho_B^{\rm net}=\rho_B - \rho_{\bar B}$ is the
net baryon density. The "$+$" and "$-$" signs are for kaons and
antikaons, respectively.

The pion potentials are related to their self-energies
$\Pi_s^{\pm0}$ according to $U_{\pi^{\pm0}} =
\Pi_s^{\pm0}/(2m_\pi)$, where $m_\pi$ is the pion mass. In
Ref.~\cite{Kai01}, the contribution of the pion-nucleon $s$-wave
interaction to the pion self-energy has been calculated up to the
two-loop order in chiral perturbation theory. In asymmetric nuclear
matter of proton density $\rho_p$ and neutron density $\rho_n$, the
resulting $\pi^-$ and $\pi^+$ self-energies are given, respectively,
by
\begin{eqnarray}\label{self}
\Pi_s^-(\rho_p,\rho_n)&=&\rho_n[T^-_{\pi N}-T^+_{\pi
N}]-\rho_p[T^-_{\pi N}+T^+_{\pi N}]
\notag\\
&&+\Pi^-_{\rm rel}(\rho_p,\rho_n)+\Pi^-_{\rm cor}(\rho_p,\rho_n)\notag\\
\Pi_s^+(\rho_p,\rho_n)&=&\Pi_s^-(\rho_n,\rho_p).
\end{eqnarray}
In the above, $T^\pm$ are the isospin-even and isospin-odd $\pi N$
$s$-wave scattering $T$-matrices, which are given by the one-loop
contribution in chiral perturbation theory and have the empirical
values $T^-_{\rm \pi N}\approx 1.847~{\rm fm}$ and $T^+_{\rm \pi
N}\approx -0.045~{\rm fm}$ extracted from the energy shift and width
of the 1$s$ level in pionic hydrogen atom; $\Pi^-_{\rm rel}$ is due
to the relativistic correction; and $\Pi^-_{\rm cor}$ is the
contribution from the two-loop order in chiral perturbation theory.
Their expressions can be found in Ref.~\cite{Kai01}. For nucleon
resonances and strange baryons in a hadronic matter, we simply
extend the above result by treating them as neutron- or proton-like
according to their isospin structure~\cite{Bal95} and light quark
numbers. Because of the $G$-parity invariance, the contributions of
antiprotons and antineutrons in the hadronic matter are similar to
those of neutrons and protons, respectively. We neglect in the
present study the pion-nucleon $p$-wave interaction~\cite{Bro75},
which is expected to reduce and enhance, respectively, the $\pi^+$
and $\pi^-$ potential difference due to the $s$-wave interaction in
the neutron-rich and proton-rich matter~\cite{Xu10}, since its
inclusion in the transport model is highly nontrivial~\cite{Xio93}.

\begin{figure}[h]
\begin{minipage}[b]{1.0\linewidth}
\includegraphics[scale=0.4]{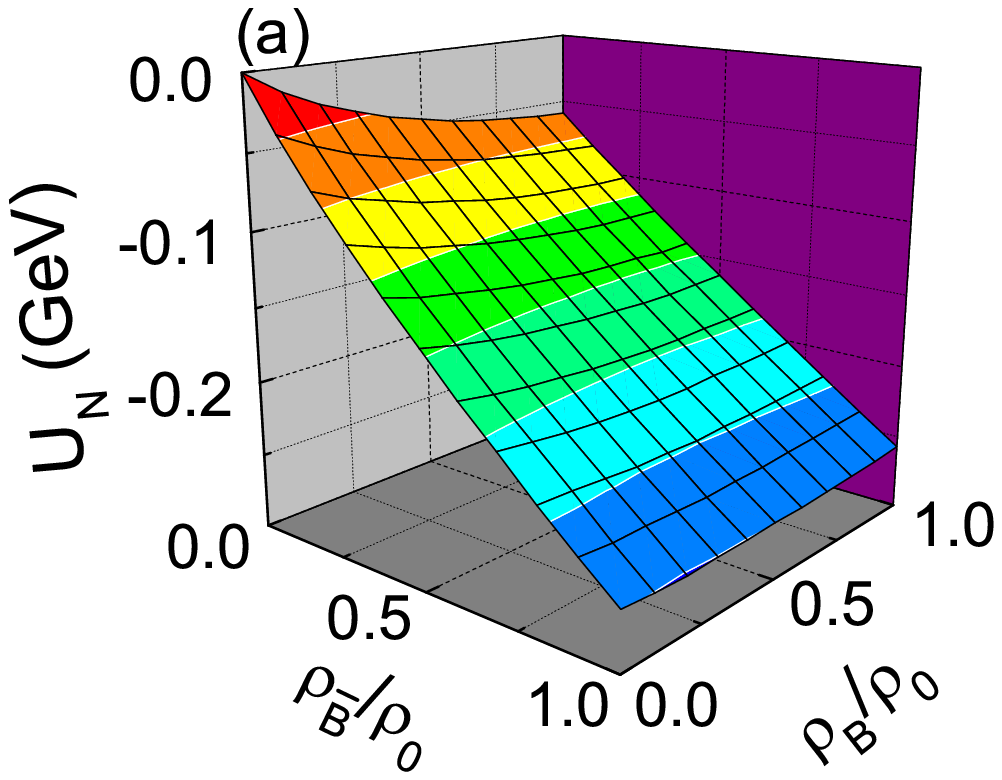}\includegraphics[scale=0.4]{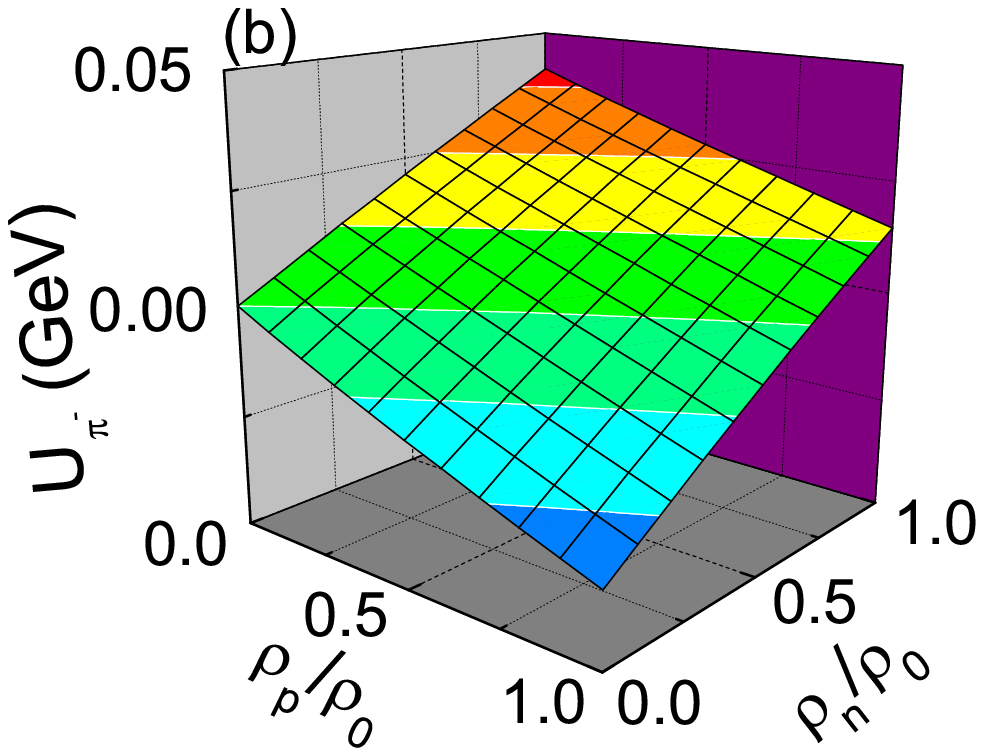}
\end{minipage}
\begin{minipage}[b]{1.0\linewidth}
\includegraphics[scale=0.4]{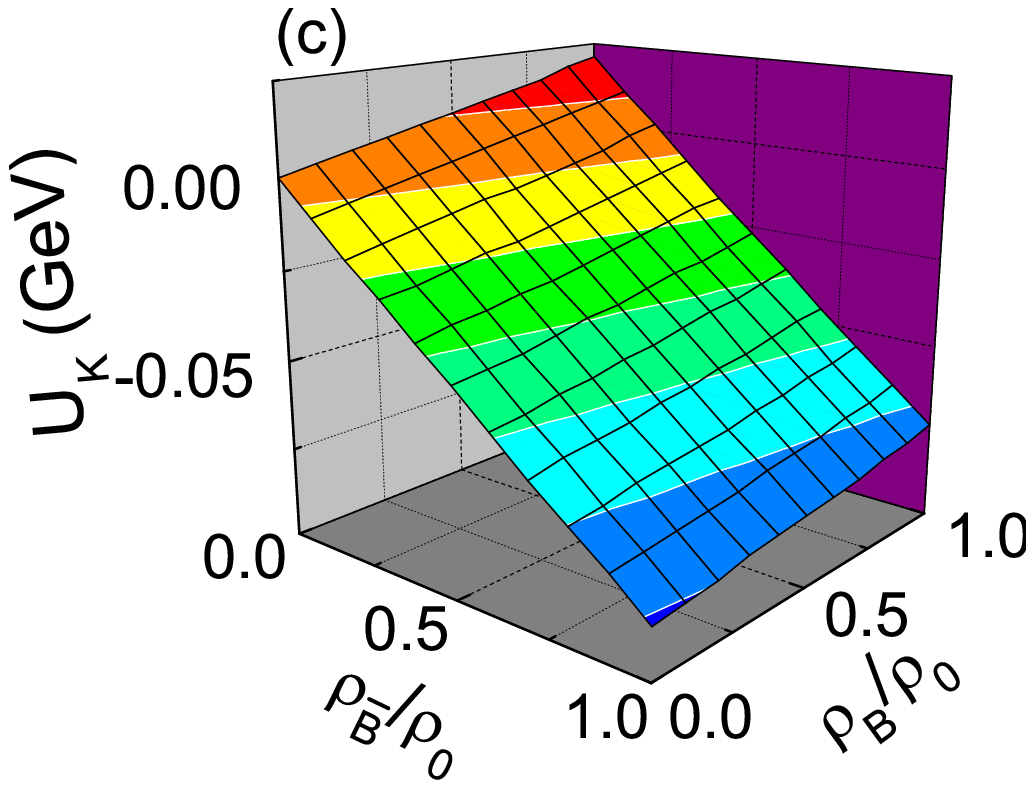}\includegraphics[scale=0.4]{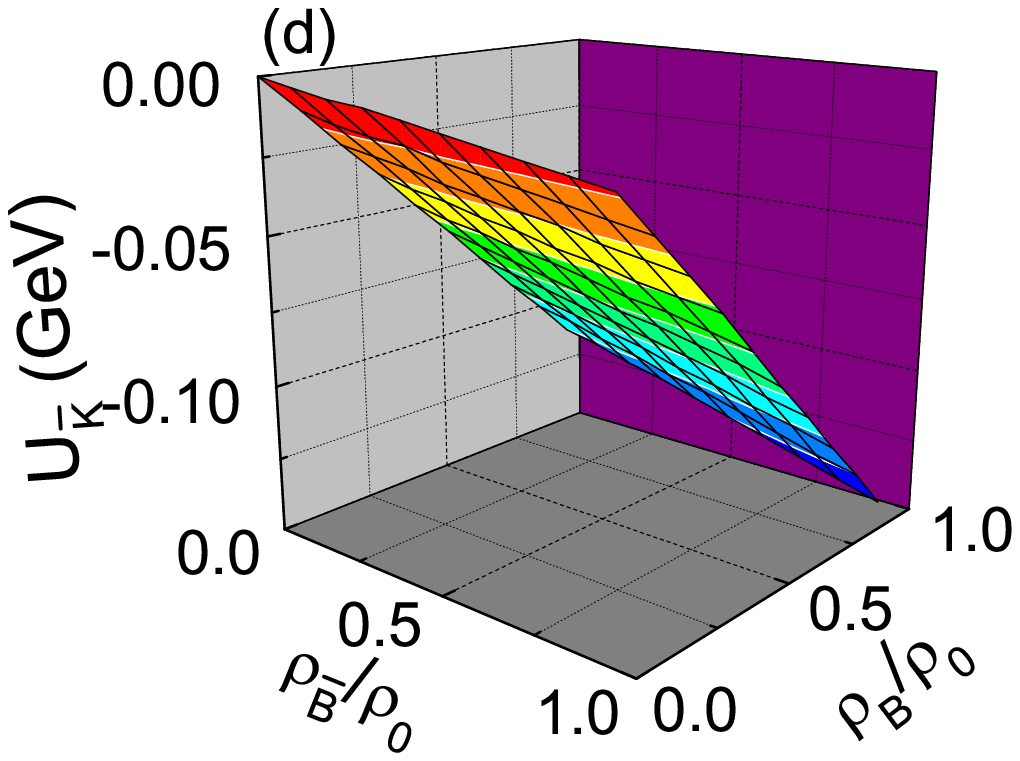}
\end{minipage}
\caption{(Color online) Mean-field potentials of $N$ (a), $K$ (c),
and ${\bar K}$ (d) at zero momentum as functions of baryon and
antibaryon densities and of $\pi^-$ (b) as a function of
neutron-like and proton-like densities.} \label{potential}
\end{figure}

In Fig.~\ref{potential}, we show the $N$, $K$, and ${\bar K}$
potentials as functions of $\rho_B$ and $\rho_{\bar B}$ and the
$\pi^-$ potential as a function of neutron-like and proton-like
densities $\rho_n$ and $\rho_p$. The ${\bar N}$ and $\pi^+$
potentials are related to those of $N$ and $\pi^-$ by $U_{\bar
N}(\rho_{\bar B},\rho_{B})= U_N(\rho_{B},\rho_{\bar B})$ and
$U_{\pi^+}(\rho_p,\rho_n)=U_{\pi^-} (\rho_n,\rho_p)$. In the absence
of antibaryons, the $N$ potential is slightly attractive while that
of ${\bar N}$ is strongly attractive, with values of about $-60$ MeV
and $-260$ MeV, respectively, at normal nuclear matter density
$\rho_0=0.16$ fm$^{-3}$. The latter is similar to that determined
from the non-linear derivative model for small antinucleon kinetic
energies~\cite{Gai11} and is also consistent with those extrapolated
from experimental data~\cite{Bar72,Pot78,Bat81,Won84,Jan86,Fri05}.
For pions in neutron-rich nuclear matter, the potential is weakly
repulsive and attractive for $\pi^-$ and $\pi^+$, respectively, and
the strength at $\rho_0$ and isospin asymmetry
$\delta=(\rho_n-\rho_p)/(\rho_n+\rho_p)=0.2$ is about $14$ MeV for
$\pi^-$ and $-1$ MeV for $\pi^+$. In antibaryon-free matter, the $K$
potential is slightly repulsive while the ${\bar K}$ potential is
deeply attractive, and their values at $\rho_0$ are about $20$ MeV
and $-120$ MeV, respectively, similar to those extracted from the
experimental data~\cite{Fri94,Bra97,Sib98,Fri99} and used in the
previous study~\cite{Son99}.

\begin{figure}[h]
\centerline{\includegraphics[scale=0.7]{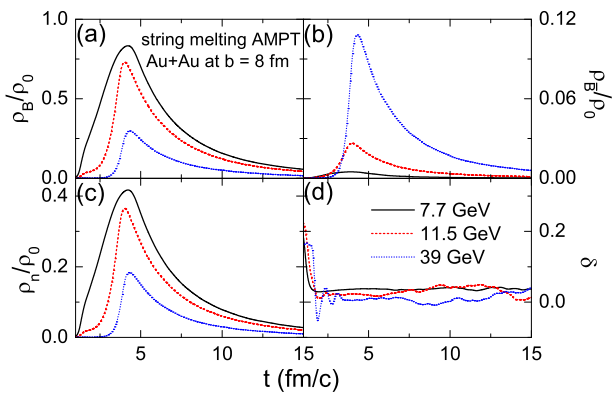}} \caption{(Color
online) Time evolution of the baryon density (a), antibaryon density
(b), neutron-like density (c), and isospin asymmetry (d) in the
central region of the hadronic phase from the AMPT model for Au+Au
collisions at $b=8$ fm and $\sqrt{s_{NN}}=7.7$, $11.5$, and $39$
GeV.} \label{den}
\end{figure}

Figure \ref{den} displays the time evolution of the baryon density
(a), antibaryon density (b), neutron-like density (c), and isospin
asymmetry (d) in the central region of the hadronic phase in Au+Au
collisions at impact parameter $b=8$ fm for the three different BES
energies of $\sqrt{s_{NN}}=7.7$, 11.5, and 39 GeV. It is seen that
the baryon density decreases while the antibaryon density increases
with increasing collision energy, resulting in a decrease of the net
baryon density with increasing collision energy. Also, the isospin
asymmetry is very small in the hadronic phase for all three energies
due to the considerable number of $\Lambda$ hyperons which do not
carry isospin and the larger number of $\pi^-$ than $\pi^+$ produced
in the collisions.

\begin{figure}[h]
\centerline{\includegraphics[scale=0.8]{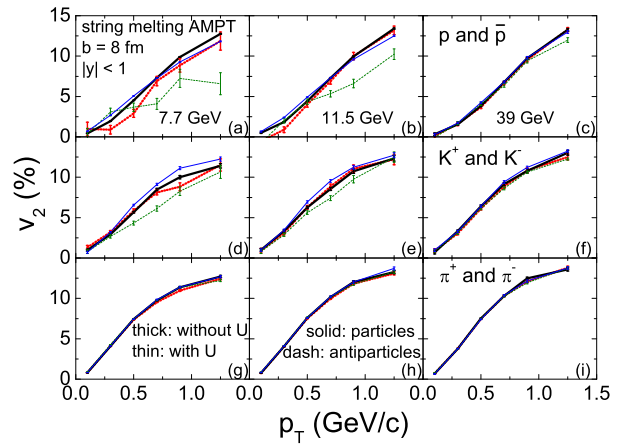}} \caption{(Color
online) Differential elliptic flows of mid-rapidity ($|y|<1$) $p$
and ${\bar p}$ [(a), (b), (c)], $K^+$ and $K^-$ [(d), (e), (f)], and
$\pi^+$ and $\pi^-$ [(g), (h), (i)] with and without hadronic
potentials $U$ in Au+Au collisions at $b=8$ fm and
$\sqrt{s_{NN}}=7.7$, $11.5$, and $39$ GeV from the string melting
AMPT model.} \label{v2pt}
\end{figure}

The differential elliptic flows of $p$, $K^+$, and $\pi^+$ as well
as their antiparticles with and without hadronic potentials at three
different BES energies from the string melting AMPT model are shown
in Fig.~\ref{v2pt}. They are calculated with respect to the
participant plane, that is $v_2=\langle\cos[2(\phi-\Psi_2)]\rangle$,
where $\phi=\rm atan2(p_y,p_x)$ is the azimuthal angle, $\Psi_2=[\rm
atan2(\langle r_p^2\sin2\phi_p\rangle,\langle
r_p^2\cos2\phi_p\rangle)+\pi]/2$ is the angle of the participant
plane, and $r_p$ and $\phi_p$ are the polar coordinates of the
participants. Without hadronic potentials the elliptic flows from
the AMPT model are similar for particles and their antiparticles.
Including hadronic potentials increases slightly the $p$ and ${\bar
p}$ elliptic flows at $p_T<0.5$ GeV/c, while reduces slightly
(strongly) the $p$ (${\bar p}$) elliptic flow at higher $p_T$,
consistent with the expectations from the relative strength of the
attractive potentials for $N$ and ${\bar N}$ shown in
Fig.~\ref{potential}. Hadronic potentials also increase slightly the
elliptic flow of $K^+$ while reduces mostly that of $K^-$, again
consistent with what is expected from the $K$ and ${\bar K}$
potentials shown in Fig.~\ref{potential}. In addition, the effect
from the potentials on the elliptic flow decreases with increasing
collision energy, which is consistent with the decreasing baryon
density and net baryon density of produced hadronic matter with
increasing collision energy shown in Fig.~\ref{den}. The difference
between the differential elliptic flows of $p$ and ${\bar p}$, and
between those of $K^+$ and $K^-$ below $\sqrt{s_{NN}}=11.5$ GeV are
qualitatively consistent with the experimental data~\cite{Moh11},
while that of $\pi^-$ and $\pi^+$ is small in all three energies due
to the small isospin asymmetries shown in Fig.~\ref{den}.

\begin{figure}[h]
\centerline{\includegraphics[scale=0.8]{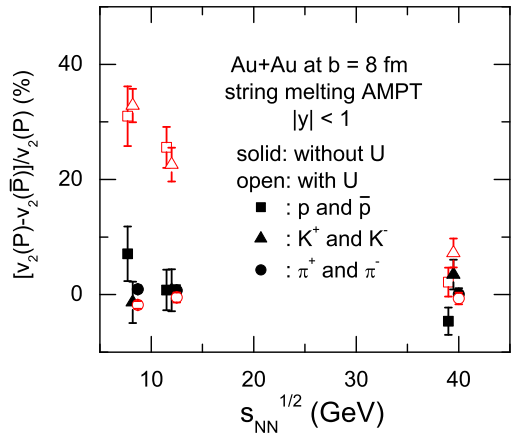}} \caption{(Color
online) Relative elliptic flow difference between $p$ and ${\bar
p}$, $K^+$ and $K^-$, and $\pi^+$ and $\pi^-$ with and without
hadronic potentials $U$ at three different BES energies from the
string melting AMPT model. Results for different species are
slightly shifted in energy to facilitate the presentation.}
\label{ratio}
\end{figure}

Our results for the relative $p_T$-integrated $v_2$ difference
between particles and their antiparticles, defined by
$[v_2(P)-v_2(\bar P)]/v_2(P)$, with and without hadronic potentials
are shown in Fig.~\ref{ratio}. These differences are very small in
the absence of hadronic potentials. Including hadronic potentials
increases the relative $v_2$ difference between $p$ and ${\bar p}$
and between $K^+$ and $K^-$ up to about $30\%$ at $7.7$ GeV and
$20\%$ at $11.5$ GeV but negligibly at $39$ GeV. These results are
qualitatively consistent with the measured values of about 63\% and
13\% at $7.7$ GeV, 44\% and 3\% at $11.5$ GeV, and 12\% and 1\% for
the relative $v_2$ difference between $p$ and ${\bar p}$ and between
$K^+$ and $K^-$, respectively~\cite{Moh11}. Similar to the
experimental data, the relative $v_2$ difference between $\pi^+$ and
$\pi^-$ is negative at all energies after including their
potentials, although ours have smaller magnitudes. We have also
found that, as seen in the experiments~\cite{Moh11}, the relative
$v_2$ difference between $\Lambda$ hyperons and $\bar \Lambda$ is
smaller than that between $p$ and ${\bar p}$, because the $\Lambda
(\bar \Lambda)$ potential is only $2/3$ of the $p (\bar p)$
potential.

To summarize, we have studied the elliptic flows of $p$, $K^+$,
$\pi^+$ and their antiparticles in heavy ion collisions at BES
energies by extending the string melting AMPT model to include their
mean-field potentials in the hadronic stage. Because of the more
attractive ${\bar p}$ than $p$ potentials, the attractive $K^-$ and
repulsive $K^+$ potentials, and the slightly attractive $\pi^+$ and
repulsive $\pi^-$ potentials in the baryon- and neutron-rich matter
formed in these collisions, smaller elliptic flows are obtained for
${\bar p}$, $K^-$, and $\pi^+$ than for $p$, $K^+$, and $\pi^-$.
Also, the difference between the elliptic flows of particles and
their antiparticle is found to decrease with increasing collision
energy as a result of decreasing baryon chemical potential of the
hadronic matter. Although our results are qualitatively consistent
with the experimental observations, they somewhat underestimated the
relative elliptic flow difference between $p$ and ${\bar p}$ as well
as that between $\pi^-$ and $\pi^+$ and overestimated that between
$K^+$ and $K^-$. In our studies, we have, however, not included
other effects that may affect the $v_2$ difference between particles
and their antiparticles. For example, we may have overestimated the
annihilation between baryons and antibaryons as this could be
screened by other particles in the hadronic matter~\cite{Kah93}.
Including the screening effect would increase the duration of the
attractive potential acting on antibaryons and thus reduces their
elliptic flow, leading therefore to an increase in the difference
between the elliptic flows of baryons and antibaryons. Also, the
different elliptic flows between particles and their antiparticles
are assumed in the present study to come entirely from the hadronic
mean-field potentials. As shown in Ref.~\cite{Plu10}, the collective
flow of partons can also be affected by their mean-field potentials
in the partonic matter. If quarks and antiquarks have different
mean-field potentials in the partonic matter, this would then lead
to different elliptic flows for particles and their antiparticles in
the initial stage of the hadronic phase after hadronization. It will
be of great interest to include in future studies these effects as
well as the effect due to different elliptic flows between produced
and transported partons~\cite{Dun11} and the chiral magnetic
effect~\cite{Bur11} in order to understand more quantitatively the
different elliptic flows between particles and their antiparticles
observed in relativistic heavy ion collisions.

This work was supported in part by the U.S. National Science
Foundation under Grants No. PHY-0758115 and No. PHY-106857, the
Welch Foundation under Grant No. A-1358, the NNSF of China under
Grant Nos. 10975097 and 11135011, Shanghai Rising-Star Program under
grant No. 11QH1401100, and "Shu Guang" project supported by Shanghai
Municipal Education Commission and Shanghai Education Development
Foundation.

\end{document}